# A New Magnetic Topological Quantum Material Candidate by Design

*Xin Gui,[a] Ivo Pletikosic,[bc] Huibo Cao,[d] Hung-Ju Tien,[e] Xitong Xu,[f] Ruidan Zhong,[b] Guangqiang Wang,[f] Tay-Rong Chang,[e] Shuang Jia,[fgh] Tonica Valla,[c] Weiwei Xie [a]\* and Robert J. Cava [b]\**

[a] Department of Chemistry, Louisiana State University, Baton Rouge, LA 70803
[b] Department of Chemistry, Princeton University, Princeton, NJ 08540
[c] Condensed Matter Physics and Materials Science, Brookhaven National Laboratory, Upton, NY, 11973
[d] Neutron Scattering Division, Oak Ridge National Laboratory, Oak Ridge, TN 37831
[e] Department of Physics, National Cheng Kung University, Tainan, Taiwan, 70101
[f] International Center for Quantum Materials, School of Physics, Peking University, Beijing, People's Republic of China 100871
[g] Collaborative Innovation Center of Quantum Matter, Beijing, People's Republic of China 100871
[h] CAS Center for Excellence in Topological Quantum Computation, University of Chinese Academy of Sciences, Beijing, People's Republic of China 100190

## *Abstract*

Magnetism, when combined with an unconventional electronic band structure, can give rise to forefront electronic properties such as the quantum anomalous Hall effect, axion electrodynamics and Majorana fermions. Here we report the characterization of high-quality crystals of $EuSn_2P_2$, a new quantum material specifically designed to engender unconventional electronic states plus magnetism. $EuSn_2P_2$ has a layered, $Bi_2Te_3$-type structure. Ferromagnetic interactions dominate the Curie-Weiss susceptibility, but a transition to antiferromagnetic ordering occurs near 30 K. Neutron diffraction reveals that this is due to two-dimensional ferromagnetic spin alignment within individual Eu layers and antiferromagnetic alignment between layers - this magnetic state surrounds the Sn-P layers at low temperatures. The bulk electrical resistivity is sensitive to the magnetism. Electronic structure calculations reveal that $EuSn_2P_2$ might be a strong topological insulator, which can be a new magnetic topological quantum material (MTQM) candidate. The calculations show that surface states should be present, and they are indeed observed by ARPES measurements.

*Introduction*

The design and discovery of new quantum materials with desired electronic and magnetic functionality, especially materials where quantum effects result in topological electronic properties, is a new challenge and one of the frontiers of research in materials chemistry and materials physics.[1] There is a crucial need for the discovery of new materials that in particular harness the topological quantum states of matter, in order to increase our understanding of these newly recognized electronic states and to realize their potential in devices.[2,3] In this context, topological insulators (TIs), which support fundamentally different surface (i.e. 'topologically protected') and bulk electronic states, are of particular interest.[4] When additional quantum phases such as superconductivity, magnetism or charge density waves are also present, new electronic properties can arise due to their interactions with the surface states.[5,6] Currently, the study of such systems in materials physics frequently focuses on heterostructures consisting of TI/magnetic multilayers.[7-10] Unfortunately, these systems cannot be widely studied due to the uncommon procedures required to make them. Thus, the design of novel magnetic topological quantum materials that are widely synthetically accessible is highly sought. Theorists have scoured the ICSD for previously reported materials that can display topological electronic properties, but the design of totally new topological quantum materials remains rare, especially for cases where the interactions of different quantum states is expected.[11-16]

Located at the Zintl border in the periodic table, the element Sn has moderate electronegativity, thus allowing it to participate in various types of bonding interactions. This provides flexibility in tuning the electronic properties of materials in which the electronic states of tin are dominant.[17] Further, being a 5$p$-based element, it is expected to display the effects of spin orbit coupling.[18] Tin may be the critical elemental constituent in the topological crystalline insulator SnTe for example, and the fabrication of Stanene thin films has been accomplished with the goal of establishing its topological electronic properties.[19,20] Thus, the inclusion of tin in quantum materials is a good chemical rule when designing quantum materials with potentially unconventional electronic functionality.

A second chemical design rule for finding novel functional materials is based on symmetry and crystal structure. Experience has shown that low dimensional materials are ideal platforms for realizing topological electronic states due to the weak interlayer interactions and subsequent partial confinement of the electronic states that is frequently present. Rhombohedral $Bi_2Se_3$[21] and $Bi_2Te_3$[22], and trigonal $Mg_3Bi_2$ are examples.[23] The presence of elements with strong spin orbit coupling in layers based on triangles is a common structural motif in topological quantum materials.

Finally, as a third design rule, the inclusion of magnetic lanthanide elements, if well placed in the crystal structure, can result in the presence of ordered magnetic states that interact relatively weakly with the other electronic states present, due to the localized nature of 4$f$ electrons.[24] With appropriately selected rare earth elements at the appropriate concentration in the appropriate crystal structure, magnetically ordered states can be found at experimentally accessible temperatures. In materials with unconventional electronic structures, this allows for the interactions between magnetic and topological states to be studied.

Although non-magnetic, the ternary $Bi_2Te_3$-type compound $SrSn_2As_2$ has been theoretically predicted to host three-dimensional Dirac states around the Fermi level.[25-27] Moreover, $NaSn_2As_2$, with the same structure, is reported to be a superconductor below 1.3 K.[28] Consisting of buckled SnAs honeycomb networks separated by $Sr^{2+}$ or $Na^+$, consideration of these compounds and the above materials design criteria inspired us to incorporate a magnetic rare earth element ($Eu^{2+}$) between Sn-based layers with a geometry based on triangles to yield both an unconventional electronic band structure and magnetic properties in a single material.

Thus, here we report a new material by design, $EuSn_2P_2$. We report its crystallographically and magnetically ordered structures, determined via single crystal X-ray and neutron diffraction. We observe a transition to an A-type antiferromagnetic state (alternating-orientation ferromagnetic layers) near 30 K, and below that temperature significant magnetoresistance. a type-II nodal-line semimetal when spin-orbit coupling is ignored. Electronic structure calculations reveal that $EuSn_2P_2$ is an antiferromagnetic topological insulator when Hubbard (U) is included. The detailed interplay of unconventional electronic states ("band inversion"), spin orbit coupling, and the surface states are presented theoretically. The presence of the surface states is confirmed by ARPES measurements. Our new quantum material is easily grown as bulk crystals, displays magnetic ordering at accessible temperatures, and displays topological surface states, thus allowing the interactions of ferromagnetism and topological surface states to be observed in a bulk material with a simple crystal structure.

## Experimental Details

**Single Crystal Growth:** To obtain single crystals of $EuSn_2P_2$, Eu chunks, red phosphorus and Sn granules were placed in an alumina crucible with the molar ratio of 1.1:2:20. The crucible was sealed into an evacuated quartz tube which was then heated to 600 °C and held for 24 hours followed by a 2-day heat treatment at 1050 °C. After that, the tube was cooled to 600 °C at a rate of 3°C/h. Excess Sn was centrifuged out after reaching 600 °C and shiny hexagonal single crystals (~1×1×0.1 mm³) were obtained.

**Structure Determination.** Multiple $EuSn_2P_2$ crystals (~60×60×5 μm³) were tested by single crystal X-ray diffraction in order to determine the crystal structure of the new material. The structure, consistent among all crystals, was determined using a Bruker Apex II diffractometer equipped with Mo radiation ($\lambda_{K\alpha}$= 0.71073 Å) at room temperature. Glycerol was used to protect the samples, which were mounted on a Kapton loop. The scan width was set to 0.5° with an exposure time of 10 seconds. Four different positions of the goniometer and detector were selected. Data acquisition was made *via* Bruker SMART software with the corrections for Lorentz and polarization effects included.[29] The direct method and full-matrix least-squares on $F^2$ procedure within the SHELXTL package were employed to solve the crystal structure.[30]

**Physical Property Measurement:** DC magnetization data were measured from 2 to 300 K using a Quantum Design Dynacool Physical Property Measurement System (PPMS), equipped with a vibrating sample magnetometer (VSM) option. The magnetic susceptibility was defined as M/H, where the applied field was 500 Oe. Field-dependent magnetization data was collected at 2 K with magnetic field up to $\mu_0 H$ = 9 T. The electronic transport measurements were carried out in a PPMS-9 cryostat using the four-terminal method between 1.8 K to 300 K under magnetic fields up to $\mu_0 H$ = 9 T. Platinum wires were attached to the samples by silver epoxy to ensure ohmic contact. The normal resistivity, $\rho_{xx}$, was measured in the basal plane of a single crystal, simply determined by dividing the observed voltage difference by the applied current, suitably normalized by the sample dimensions. The Hall resistance, $\rho_{yx}$, was similarly determined by applying the magnetic field perpendicular to a single crystal basal plane and measuring the voltage within the plane perpendicular to the direction of the applied current (again suitably normalized by the sample size).

**Electron Paramagnetic Resonance (EPR):** A single crystal of EuSn$_2$P$_2$ was stored in the glovebox before EPR measurements, and then placed in a closed sample tube. The sample tube was put into the EPR cavity. The experiment was carried out on a Bruker ER 300D spectrometer interfaced to a Bruker 1600 computer for data storage and manipulations. The spectrum was collected by sweeping the magnetic field from 0 to 5000 Gauss.

**Single Crystal Neutron Diffraction:** Single crystal neutron diffraction was performed on the Four-Circle Diffractometer (HB-3A) at the High Flux Isotope Reactor (HFIR) at Oak Ridge National Laboratory (ORNL). Due to the large neutron absorption coefficient of Eu, a relatively small single crystal (~0.8×0.4×0.02 mm$^3$) was selected and mounted on an aluminum pin. The measurement was carried out between 4 K and 50 K with an incident neutron wavelength of 1.550 Å from a bent perfect Si-220 monochromator. Absorption corrections were applied using the PLATON software.[31,32] The nuclear and magnetic structure refinements were carried out using the FULLPROF refinement Suite.[33]

**Electronic Structure Calculations:** The bulk electronic structure calculations for EuSn$_2$P$_2$ were performed by using the projector augmented wave method as implemented in the VASP package within the generalized gradient approximation (GGA) scheme.[34,35] The experimental structural parameters were employed. Spin-orbit coupling was included self-consistently in some of the calculations for comparison to the expected electronic structure with SOC absent. Spin polarization using GGA+U (U= 4eV) was employed for the magnetic models investigated.[36] How *f* electron states are treated in DFT calculations is of interest because when considered as valence electrons then they are always found at the Fermi energy for partly filled *f* orbitals, which, except in very rare cases such as for heavy fermion materials[37,38], is not a realistic picture for their relative position in energy - thus in our calculations, the Eu *f* electron states are treated as core rather than valence states. A Monkhorst-Pack 11×11×7 *k*-point mesh was used for the *k*-space integrations.[39] With these settings, the calculated total energy converged to less than 0.1 meV per atom. For the surface spectral weight calculations, a semi-infinite Green's function method, terminated by (001) surfaces, was employed. Calculations were made for Sn, P and Eu-terminated (001) surfaces. The median line of the Sn bilayer is selected as the Sn-terminated surface.

**Experimental Band Structure Characterization:** Angle-resolved photoelectron spectroscopy (ARPES) measurements were carried out at a temperature of 10 K on the (001) surfaces of EuSn$_2$P$_2$ crystals cleaved in a $10^{-9}$ Pa vacuum. Linearly (*s* & *p*) polarized light with photon energies in the range 50-120 eV was used. Experiments were performed at the Electron Spectro-Microscopy (ESM) beamline of the National Synchrotron Light Source (NSLS-II). The photoemitted electrons were collected from a ±15° cone by a Scienta DA30 and analyzed with angular and energy resolutions of < 0.2° and 15 meV respectively. Core level spectra using 300 eV photons showed presence of no elements other than Sn, Eu, and P in the material. Potassium adsorption, commonly used to fill a material's bands with additional electrons to increase the Fermi Level at the surface of a sample, resulted in surface deterioration in the case of $EuSn_2P_2$ and no noticeable Fermi level shift.

## *Results and Discussion*

**Crystal Structure:** The single crystals of $EuSn_2P_2$ are hexagonal plates, as shown in Fig. 1(A), with the rhombohedral *c*-axis perpendicular to the plates (thus the plates are (001) or "basal" planes.) The results from single crystal X-ray diffraction reveal that $EuSn_2P_2$ crystallizes in a rhombohedral lattice with the space group *R*-3*m*, in analogy to $Bi_2Te_3$, which is a well-known topological insulator. As shown in Fig. 1(B), each Eu atom, surrounded by six P atoms, forms a trigonal prism. The Eu layers are separated by two $Sn@P_3$ triangular pyramid layers connected by weak Sn-Sn bonds. An inversion center exists in the center of the unit cell. The layered features of $EuSn_2P_2$, and its in-plane and honeycomb Sn structure viewed along *c*-axis are presented in Fig. 1(C) and (D), respectively. The crystallographic data, including atomic positions, site occupancies, and equivalent isotropic thermal displacement parameters are listed in Tables 1 and 2. The refined anisotropic displacement parameters are shown in Table S1.

**Magnetic Characterization:** In order to get insight into the magnetic properties of $EuSn_2P_2$, we performed field-dependent magnetization measurements at 2K with applied magnetic fields parallel and perpendicular to the *c*-axis. The results are shown in Fig. 2(A). The field-dependent behavior is characteristic of that of a ferromagnetic material – the magnetic moment of $EuSn_2P_2$ saturates at ~2.8 T for applied fields perpendicular to the *c*-axis and ~4.1 T for fields parallel to the *c*-axis. Accordingly, a strong anisotropy of magnetization is observed, with the easy axis in the basal plane, perpendicular to the *c*-axis. The saturated magnetizations are ~6.0$\mu_B$/Eu for both orientations.

The temperature-dependent magnetic susceptibility was measured under an external field of 500 Oe from 2 to 300 K. A transition to an antiferromagnetic state (i.e. $T_N$) is seen at around 30 K. The larger magnetic susceptibility in the perpendicular field also indicates the anisotropic ordered magnetism in EuSn$_2$P$_2$; i.e. that ferromagnetic ordering is preferred within the *ab*-plane rather than along the *c*-axis. At temperatures higher than 30K, classic Curie-Weiss behavior is observed.[40] The effective moment and Curie-Weiss (CW) temperatures $\Theta_\perp$ and $\Theta_{//}$ were fitted from 150 K to 300 K by using Curie-Weiss formula $\chi = \chi_0 + \frac{N_A}{3k_B}\frac{\mu_{eff}^2}{(T-\theta_{CW})}$ where $\chi_0$ is the temperature independent part of the magnetic susceptibility, $N_A$ is Avogadro's number, $k_B$ is the Boltzmann constant, $\mu_{eff}$ is the effective moment and $\Theta_{CW}$ is the Curie-Weiss temperature. The Curie-Weiss temperatures in perpendicular and parallel fields are 31.8(1) K and 31.6(1) K, within error of each other, while the average effective moment is 7.91(3) $\mu_B$/Eu. The magnetic characteristics above $T_N$ are essentially isotropic.

**Electron Paramagnetic Resonance (EPR):** The magnetism observed is consistent with the presence of magnetic Eu$^{2+}$ in our material, but to further confirm the electronic state of Eu in EuSn$_2$P$_2$, we performed electron paramagnetic resonance measurements. The obtained data is shown in Fig. 3 as the first derivative of the absorption data. We observe a sin-function-like strong broad curve when the applied magnetic field is swept to 2400 G, reaching its maximum around 3200 G. The intensity subsequently drops to zero at 3530 G, which indicates the applied field of the strongest absorption, and continuously becomes more negative, to ~ -6000 (in arbitrary units) at ~ 3800 Oe, returning to zero intensity beyond 5000 Oe. The asymmetric line shape is as is usually observed in metals due to the skin effect caused by the small size of the single crystal used compared with the skin depth. According to the principal EPR equation: $\Delta E = h\nu = g\mu_B B_0$ where $\Delta E$ is the splitting energy, *h* is Planck's constant, $\nu$ is the frequency of the microwave used, which is 9481 MHz, *g* is the Landé *g*-factor, $\mu_B$ is the Bohr magneton and $B_0$ is the applied magnetic field of the strongest absorption, we obtain the Landé *g*-factor of 1.9184. Based on $\mu_{eff} = g\sqrt{J(J+1)}\mu_B$ and the *g*-factor obtained from the EPR data, we obtain $\mu_{eff}$ = 7.61 $\mu_B$, which is within error of the effective moment obtained from the temperature dependent magnetic susceptibility (7.91 $\mu_B$).[41] Therefore, we find that for Eu in EuSn$_2$P$_2$, *J* = 7/2 and that the Eu ground state is $^8S_{7/2}$, i.e. that Eu$^{2+}$ is the primary valence state for Eu.

**Magnetic Structure:** To better understand the magnetic properties of our new quantum material, we performed single crystal neutron diffraction on EuSn$_2$P$_2$. Antiferromagnetic order onsets at 30 K, as indicted by observing extra Bragg peaks with a propagation vector of (0 0 1.5). The ordering parameter was measured at (0 0 4.5) in reciprocal space (see Fig. 4A) and indicates that $T_N \sim 30$ K by determination of the intersection of a linear fitting regime (30-50 K) and a two-term power law fit at low temperatures with $y = \begin{cases} A|x - x_c|^{pl}, x < x_c \\ A|x - x_c|^{pu}, x > x_c \end{cases}$ where $A$, $pl$, $pu$ and $x_c$ are all constants.[33] The magnetic structure is determined by modeling the magnetic reflections collected at 4.5 K. The lattice parameters ($a$=4.069(2) Å, $c$=25.935(9) Å) at 4.5 K are smaller than those determined by single crystal X-ray diffraction ($a$=4.097(1) Å, $c$=26.162(5) Å) at room temperature due to thermal contraction. Refining the nuclear reflections collected at 4.5 K yields $R_F$=9.15% and $R_F^2$=17.69%. The magnetic structure refinement reaches the best fit ($R_F$=6.71% and $R_F^2$=12.7%) with the magnetic moment of 5.65 (3) $\mu_B$/Eu oriented in the $ab$-plane at 4.5 K (see Fig. 4B), consistent with the field-dependent magnetization measurements. As shown in Fig. 4B, the result is ferromagnetic Eu layers antiferromagnetically coupled with their nearest neighbor layers leading to A-type antiferromagnetism for $EuSn_2P_2$. The insert of Fig. 4A illustrates the neutron diffraction intensity maps at both 30 K and 4.5 K; showing the magnetic peak at (0 0 4.5).

**Charge Transport:** The transport of electrical charge in materials where interactions between different states may be present is of interest. The temperature-dependent resistivity for a single crystal of $EuSn_2P_2$, measured in the (001) plane, is thus shown in Fig. 5(A). Decreasing resistivity with decreasing temperature, i.e. metallic behavior, is observed. The charge transport is sensitive to the magnetic state of the system: $\rho_{xx}$ (the resistivity in the $x$ direction with the electric field applied along the $x$ direction; $x$ and $y$ directions are in the plane of the layers, while $z$ is perpendicular to the layers) shows a drop below the Neel temperature ($\sim 30$ K). The in-plane resistivity $\rho_{xx}$ increases slightly at temperatures above the magnetic ordering temperature, starting near 58 K. Also, as seen in Fig. 5(B), a linear relation between the Hall resistivity $\rho_{yx}$ (field is applied along $z$, electric field along $x$ and resistivity measured along $y$) and applied magnetic field H is seen at various temperatures. From this data we are able to calculate the concentration of the dominant carriers ($n$), which are holes, by using $n$=1/($eR_H$) (where $R_H$ is the Hall coefficient, obtained from the slopes of Fig. 5(B) and $e$ is the electron charge.)[41] We find that at 2 K $n$ is about $2.9 \times 10^{20}$ cm$^{-3}$, a relatively large hole concentration. The slope of the $\rho_{yx}$ (H) curve decreases somewhat with increasing temperature and thus the carrier concentration is increasing with increasing temperature.

Uncommon behavior for the magnetoresistance is observed. (Fig. 5(C)) At low temperatures, the material's resistivity $\rho_{xx}$ shows a dome-like feature when magnetic field is applied perpendicular to the basal plane (i.e. along *z*). At 2 K, it decreases by about 9.4 µΩ cm when the field increased from 0 to about 4.1 T, this latter field being where M vs H saturates at the same temperature. (Fig. 2(A)) This behavior is consistent with the anomalous magnetoresistance seen in spin flop systems. The field dependent resistivity is not quenched until temperatures above 40 K, above the magnetic ordering temperature - the magnetoresistance becomes more normal for temperatures of 70 K and above. Measurements on a second sample suggest that the magnetoresistance in a direction perpendicular to the plane is positive and significantly larger than what is observed in-plane (see Fig. S1). Further, the out-of-plane resistivity does not appear to saturate by 9 T and may continue to rise with larger applied fields. Although the in-plane magnetoresistance is reminiscent of that observed for WTe$_2$, the change of resistivity in field is orders of magnitude lower than is seen there, and magnetic moments (from the Eu) are actually present in the current case, while they are absent for WTe$_2$ [42,43]. Layered ferromagnetic semimetallic (see below) systems like EuSn$_2$P$_2$ are relatively rare, and thus detailed study will be required to fully understand the magnetoresistance observed in Fig. 5(D).

**Calculated Bulk and Surface Electronic Structures:** The results of first-principles electronic structure calculations for EuSn$_2$P$_2$ using the Generalized Gradient approximation (GGA), GGA plus correlation parameter U (GGA+U), and GGA+U with spin-orbit coupling (SOC) are compared in Fig. 6. The calculations reveal a semi-metallic band structure for this material, with a continuous gap throughout the Brillouin zone, similar in character to what is observed for WTe$_2$.[42,43] The electrons in the 4*f* orbitals of the Eu atoms are experimentally observed (see below) as localized in the range of 1.0-2.0 eV below the Fermi level (E$_F$), and thus the GGA+U+SOC calculations appear to be the most representative of the material.

The calculations show that P 3*p* orbitals dominate the energy bands just below E$_F$, while Sn 5p orbitals are found below -5 eV and just above E$_F$ (Fig.6(E)). Like is the case for strong topological insulators, Fig. 6E shows that in the vicinity of the Fermi Energy there are metal states mixed in in significant proportion at the top of the valence band and P states mixed in in significant proportion at the bottom of the conducting band, i.e. there is an inversion of the conventional type of band structure, where this kind of mixing of states above and below E$_F$ is negligible. When the effects of spin polarization are included, the total energies of the ferromagnetic (FM) and anti-ferromagnetic (AFM) models are the lowest, and indicate that the observed antiferromagnetism is energy favored. The magnetic moments, calculated to be about ~6.7 $\mu_B$, are solely on the Eu atoms, consistent with what is expected in a naive picture for this material, where Eu is the only magnetic atom. The electronic structure of the material, calculated for the antiferromagnetically ordered system, shows that the bands near the Fermi level are hybridized among electrons from the *p* orbitals on the Sn and P atoms. The GGA and GGA+U calculations result in a semi-metallic ground state, with no band gap, with Fermi surfaces around the Γ and A points in the Brillouin Zone (BZ) (Fig.6(A,C)).[23] When SOC is included in the calculations, a very small (a few meV) band gap is present (Fig.6(B,D)).

As was designed, EuSn$_2$P$_2$ exhibits interesting surface states. To understand the surface states of EuSn$_2$P$_2$, surface spectral weight simulations consistent with the practice used in the topological materials field were employed. (The first BZ of the rhombohedral lattice is shown in Fig. S2) The (001) surface is our material's clear cleavage plane. The (001) surface may be terminated by Eu, P, or Sn atomic layers; surface states are visible for all three possible (001) terminations in the calculations. Consistent with the full bulk DFT calculations, the electronic behavior near $E_F$ in these calculations is predominantly determined by the dispersion of hybridized bands from Sn and P orbitals. For Sn-terminated and Eu-terminated surfaces, comparison of the bulk and surface calculations (Fig. 7) shows that surface states are located in the bulk energy gap around the Γ point. The energy dispersion of these states is very similar to the Dirac surface states of strong TIs (inset of Fig.7(D, E, F)). The surface states are more well-separated in *k* for the P-termination case (Fig. 7E) than they are for the Sn and Eu termination cases. This should make them more observable experimentally than for the case of WTe$_2$, where they are strongly confined in both energy and wavevector.

It is well-known the time-reversal topological insulators can be described by $Z_2$ invariant. Generally, $Z_2$ is undefined in time-reversal symmetry (TRS) breaking system such as ferromagnetic phase. Nonetheless, recent works proposed that the $Z_2$ invariant can be expanded to classify the topological states in magnetic system if the system possesses specific magnetic configuration, for instance, antiferromagnetic state, in which the combined symmetry S=Θ$T_{1/2}$ is preserved, where Θ is time-reversal operator and $T_{1/2}$ is lattice translational symmetry of the crystal that is broken by the antiferromagnetic order [44,45]. Under this definition, we calculate the $Z_2$ invariant by the Wilson loop method shown in Fig. 8.[46] Our calculation shows an open curve traversing the entire Brillouin zone in the time-reversal invariant plane $k_z = 0$, indicating $Z_2$ invariant equal 1, which demonstrate $EuSn_2P_2$ is indeed an antiferromagnetic topological insulator (AFM-TI). The surface state fermi contours, calculated by way of the semi-infinite Green's function,[47] are shown in Fig. 9. The calculated iso-energy band contours at E = 0, -0.2, and -0.3 eV (below the calculated Fermi energy) are shown for the three possible (001) surface terminations. (The energies correspond to those of the surface states on the Sn terminated surface.[48]) There is a prominent feature that is seen for the surface band contour in addition to the bulk states directly around the Γ point in the BZ.: A six-fold symmetry loop. The loop is small on the Sn-terminated and Eu-terminated surfaces, but is distinctly seen separated from the bulk states in *k* on the P-terminated surfaces.

**Experimentally Observed Surface Electronic Structures:** The ARPES spectra, recorded in $EuSn_2P_2$'s antiferromagnetically ordered state at 10 K, shown in Figs. 10 and 11, reveal the energies of the occupied electronic bands around the center Γ of the (001) surface projection of the bulk BZ. The outer (meaning the band with *k* vector furthest from the center of the surface BZ) band, designated as σ, disperses linearly at about -0.5 10^6 m/s away from the Fermi level ($E_F$) and appears as a sharp, intense spectral feature at all incident photon energies. This is in contrast with all of the other observed electronic bands and reflects the fact that this is a surface state. Due to the non-conservation of perpendicular-to-plane momenta of the ARPES method, a range of perpendicular momenta is observed in the ARPES spectra of the bulk bands, making them appear as filled-in regions in ARPES maps, distinct from the surface band where the absence of perpendicular dispersion results in a sharp feature. The most notable bulk band is the heart-shaped band, labeled as γ, in the center of the surface BZ, which disperses upward in energy starting from Gamma to about 80 meV below $E_F$ and turning again down. The broad bright feature showing little dispersion in the range of energies between -1 and -2 eV (below $E_F$) is from the localized Eu *f* states.

As is often the case with surface states, the surface band σ is found relatively close in *k* space to one of the bulk bands, labeled as α. Additional bulk bands below $E_F$, labeled as β and δ, are also observed. The constant energy maps, Fig. 11, taken at two different incident photon energies (hence, at different $k_z$ momenta), show the two-dimensional momentum space surface Fermi surfaces: the hexagonal-loop contour of the surface band σ encloses the contours formed by the bulk bands α, β, and γ. These features greatly resemble the -300 meV band contours calculated for the P termination of EuSn$_2$P$_2$, Figs. 9B, E, and H, although there is less warping towards the corners of the BZ for the surface band than is calculated.

Momentum space mapping that included a few neighboring surface BZs revealed no bands in the vicinity of the Fermi level other than those shown around Γ in Fig. 11. While three other crystal terminations (Eu, Sn single and double layer) cannot be excluded in this material without further theoretical and experimental studies, all our data indicate that the (001) crystal surfaces are indeed P terminated: the calculations for the Eu and Sn terminations show surface states much closer to the center of the BZ. Moreover, our core level spectra (Fig. 12) show that the P 2$s$ levels occur at two energies, indicative of two starkly different chemical environments (thus bulk vs. surface) for the P atoms. By comparing the band dispersion observed in the ARPES spectra and the DFT calculations, we conclude that the cleaved crystals are some 220 meV hole doped. This is consistent with the large $p$-type carrier concentration observed in the Hall measurements.

We note that the calculations show the bottom of six elliptical electron pockets around the M point of the BZ, which are slightly above the chemical potential (about –220 meV compared to the calculated $E_F$) observed in the ARPES experiments. This explains why no trace of those calculated bands is seen in ARPES (which sees only filled states) and further gives a hint for why the results of potassium adsorption measurements, expected to electron-dope the surface, resulted in no noticeable band shifts - the flat bottom of an electron pocket is associated with a high density of states, and all the states have to be filled for the shifts in other bands to be observed. We also note that our ARPES measurements at varying incident photon energies show remarkably little difference in the band structure of EuSn$_2$P$_2$ at different perpendicular momenta $k_z$. The bands - most notably the surface state σ, Figs 10, and 11 show little dispersion perpendicular to the layers. The three-dimensional Fermi surface thus consists of hexagonally warped cylinders formed by extruding the contours of the bulk bands α and β along the crystal axis $c$ (Fig. 11 B).

## *Summary and Conclusions*

The novel quantum material EuSn$_2$P$_2$ was designed based on chemical criteria to display both magnetism and electronic surface states. The hypothesized material was discovered, and crystals were grown using the Sn-flux method. Its crystal structure and physical properties were characterized. Its crystal structure consists of hexagonal Eu, and P layers surrounding an Sn bi-layer. It has a positive Curie-Weiss temperature of $\theta_{CW} = 31(1)$ K, showing the dominance of ferromagnetic spin correlations at high temperatures, but it orders antiferromagnetically near 30 K. Its ordered magnetic structure, refined from neutron diffraction, shows antiferromagnetically coupled in-plane-ferromagnetic $Eu^{2+}$ layers. The temperature-dependent resistance of $EuSn_2P_2$ is weakly metallic in character, and an anomalous magnetoresistance is found below the magnetic ordering temperature and below applied magnetic fields of 4 T, where the magnetization saturates with applied magnetic field.

Electronic structure calculations support magnetic ordering as the energetically stable state of the system at low temperatures, and further confirm that the observed magnetic moments solely arise from the $Eu^{2+}$ present. Calculations including both SOC and correlation parameter U best describe the positions of the Eu $f$ states between 1 and 2 eV below $E_F$, and further show that $EuSn_2P_2$ is an antiferromagnetic topological insulator (the energies of the valence band and conduction bands overlap but for electrons of different wave vectors in the Brillouin Zone.) The bulk band structure is unconventional in that metal states contribute significantly just below the Fermi energy while non-metal states contribute significantly just above the Fermi energy. The calculations also predict the presence of Dirac-like surface states within the bulk energy gap, both above and below the Fermi energy. The experimental ARPES study shows that surface states are indeed present in the antiferromagnetically ordered state at 10 K but differ in a minor way from the calculations for a P terminated (001) surface. Although we have observed much about the basic properties of this new quantum material, more detailed work on the properties from the physics perspective is required to determine whether it is unique in that it shows both antiferromagnetism and surface states at low temperatures. The accessible temperatures involved, the simple crystal structure, the ability to fabricate bulk crystals and the combination of properties make the quantum material $EuSn_2P_2$ an ideal platform for the detailed study of the interplay of topological electronic states with magnetism.

## *Author Information*

Corresponding Author: weiweix@lsu.edu; rcava@princeton.edu

Notes: The authors declare no competing financial interest.

*Acknowledgements*

The work at Princeton was supported by the ARO MURI on Topological Insulators, grant W911NF1210461. W.X. at LSU was supported by a Beckman Young Investigator award. X.G. at LSU is supported by the National Science Foundation under NSF-OIA-1832967. H.B.C. acknowledges support of US DOE BES Early Career Award KC0402010 under Contract DE-AC05-00OR22725. This research used resources at the High Flux Isotope Reactor, a DOE Office of Science User Facility operated by the Oak Ridge National Laboratory. This research also used resources of the National Synchrotron Light Source II, a U.S. Department of Energy (DOE) Office of Science User Facility operated for the DOE Office of Science by Brookhaven National Laboratory under Contract No. DE-SC0012704. IP thanks J. Jiang, E. Vescovo, and K. Kaznatcheev for their help with setting up the experiment at the 21-ID-1 (ESM-ARPES) beamline. T.-R.C. was supported from Young Scholar Fellowship Program by Ministry of Science and Technology (MOST) in Taiwan, under MOST Grant for the Columbus Program MOST107-2636-M-006-004, National Cheng Kung University, Taiwan, and National Center for Theoretical Sciences (NCTS), Taiwan. W.X. and X.G. thank D. Vineyard for helping with electron paramagnetic resonance measurement.

# References

(1) Awschalom, D. D.; Bassett, L. C.; Dzurak, A. S.; Hu, E. L.; Petta, J. R. Quantum Spintronics: Engineering and Manipulating Atom-Like Spins in Semiconductors. *Science* **2013**, *339*, 1174–1179.

(2) Cirac, J. I.; Zoller, P. Goals and Opportunities in Quantum Simulation. *Nat. Phys.* **2012**, *8*, 264–266.

(3) Ladd, T. D.; Jelezko, F.; Laflamme, R.; Nakamura, Y.; Monroe, C.; O'Brien, J. L. Quantum Computers. *Nature* **2010**, *464*, 45–53.

(4) Moore, J. E. The Birth of Topological Insulators. *Nature* **2010**, *464*, 194–198.

(5) Sato, M.; Ando, Y. Topological Superconductors: A Review. *Rep. Prog. Phys.* **2017**, *80* (7), 076501.

(6) Li, R.; Wang, J.; Qi, X.-L.; Zhang, S.-C. Dynamical Axion Field in Topological Magnetic Insulators. *Nat. Phys.* **2010**, *6*, 284–288.

(7) Chang, C.-Z.; Zhao, W.; Kim, D. Y.; Zhang, H.; Assaf, B. A.; Heiman, D.; Zhang, S.-C.; Liu, C.; Chan, M. H. W.; Moodera, J. S. High-Precision Realization of Robust Quantum Anomalous Hall State in a Hard Ferromagnetic Topological Insulator. *Nat. Mater.* **2015**, *14*, 473–477.

(8) Chang, C.-Z.; Zhang, J.; Feng, X.; Shen, J.; Zhang, Z.; Guo, M.; Li, K.; Ou, Y.; Wei, P.; Wang, L.-L.; et al. Experimental Observation of the Quantum Anomalous Hall Effect in a Magnetic Topological Insulator. *Science* **2013**, *340*, 167–170.

(9) Katmis, F.; Lauter, V.; Nogueira, F. S.; Assaf, B. A.; Jamer, M. E.; Wei, P.; Satpati, B.; Freeland, J. W.; Eremin, I.; Heiman, D.; et al. A High-Temperature Ferromagnetic Topological Insulating Phase by Proximity Coupling. *Nature* **2016**, *533*, 513–516.

(10) Mogi, M.; Kawamura, M.; Yoshimi, R.; Tsukazaki, A.; Kozuka, Y.; Shirakawa, N.; Takahashi, K. S.; Kawasaki, M.; Tokura, Y. A Magnetic Heterostructure of Topological Insulators as a Candidate for an Axion Insulator. *Nat. Mater.* **2017**, *16*, 516–521.

(11) Slager, R.-J.; Mesaros, A.; Juričić, V.; Zaanen, J. The Space Group Classification of Topological Band-Insulators. *Nat. Phys.* **2013**, *9*, 98–102.

(12) Yang, K.; Setyawan, W.; Wang, S.; Buongiorno Nardelli, M.; Curtarolo, S. A Search Model for Topological Insulators with High-Throughput Robustness Descriptors. *Nat. Mater.* **2012**, *11*, 614–619.

(13) Vergniory, M. G.; Elcoro, L.; Felser, C.; Bernevig, B. A.; Wang, Z. The (High Quality) Topological Materials In The World. *arXiv:1807.10271 [cond-mat]* **2018**.

(14) Po, H. C.; Vishwanath, A.; Watanabe, H. Symmetry-Based Indicators of Band Topology in the 230 Space Groups. *Nat. Commun.* **2017**, *8* (1), 50.

(15) Tang, F.; Po, H. C.; Vishwanath, A.; Wan, X. Towards Ideal Topological Materials: Comprehensive Database Searches Using Symmetry Indicators. *arXiv:1807.09744 [cond-mat]* **2018**.

(16) Bradlyn, B.; Elcoro, L.; Cano, J.; Vergniory, M. G.; Wang, Z.; Felser, C.; Aroyo, M. I.; Bernevig, B. A. Topological Quantum Chemistry. *Nature* **2017**, *547*, 298–305.

(17) Nesper, R. The Zintl-Klemm Concept – A Historical Survey. *Z. Anorg. Allg. Chem.* **2014**, *640* (14), 2639–2648.

(18) Gui, X.; Sobczak, Z.; Chang, T.-R.; Xu, X.; Huang, A.; Jia, S.; Jeng, H.-T.; Klimczuk, T.; Xie, W. Superconducting SrSnP with Strong Sn–P Antibonding Interaction: Is the Sn Atom Single or Mixed Valent? *Chem. Mater.* **2018**, *30*, 6005–6013.

(19) Hsieh, T. H.; Lin, H.; Liu, J.; Duan, W.; Bansil, A.; Fu, L. Topological Crystalline Insulators in the SnTe Material Class. *Nat. Commun.* **2012**, *3*, 982.

(20) Deng, J.; Xia, B.; Ma, X.; Chen, H.; Shan, H.; Zhai, X.; Li, B.; Zhao, A.; Xu, Y.; Duan, W.; et al. Epitaxial Growth of Ultraflat Stanene with Topological Band Inversion. *Nat. Mater.* **2018**, *17*, 1081.

(21) Checkelsky, J. G.; Hor, Y. S.; Cava, R. J.; Ong, N. P. Bulk Band Gap and Surface State Conduction Observed in Voltage-Tuned Crystals of the Topological Insulator $Bi_2Se_3$. *Phys. Rev. Lett.* **2011**, *106*, 196801.

(22) Zhang, H.; Liu, C.-X.; Qi, X.-L.; Dai, X.; Fang, Z.; Zhang, S.-C. Topological Insulators in $Bi_2Se_3$, $Bi_2Te_3$ and $Sb_2Te_3$ with a Single Dirac Cone on the Surface. *Nat. Phys.* **2009**, *5*, 438–442.

(23) Chang, T.-R.; Pletikosic, I.; Kong, T.; Bian, G.; Huang, A.; Denlinger, J.; Kushwaha, S. K.; Sinkovic, B.; Jeng, H.-T.; Valla, T.; et al. Realization of a Type-II Nodal-Line Semimetal in $Mg_3Bi_2$. *Adv. Sci.* **2019**, *6*, 1800897.

(24) Borisenko, S.; Evtushinsky, D.; Gibson, Q.; Yaresko, A.; Kim, T.; Ali, M. N.; Buechner, B.; Hoesch, M.; Cava, R. J. Time-Reversal Symmetry Breaking Type-II Weyl State in $YbMnBi_2$. *arXiv:1507.04847 [cond-mat]* **2015**.

(25) Rong, L.-Y.; Ma, J.-Z.; Nie, S.-M.; Lin, Z.-P.; Li, Z.-L.; Fu, B.-B.; Kong, L.-Y.; Zhang, X.-Z.; Huang, Y.-B.; Weng, H.-M.; et al. Electronic Structure of $SrSn_2As_2$ near the Topological Critical Point. *Sci. Rep.* **2017**, *7* (1), 6133.

(26) Heremans, J. P.; Cava, R. J.; Samarth, N. Tetradymites as Thermoelectrics and Topological Insulators. *Nat. Rev. Mater.* **2017**, *2* (10), 17049.

(27) Cava, R. J.; Ji, H.; Fuccillo, M. K.; Gibson, Q. D.; Hor, Y. S. Crystal Structure and Chemistry of Topological Insulators. *J. Mater. Chem. C* **2013**, *1*, 3176–3189.

(28) Goto, Y.; Yamada, A.; Matsuda, T. D.; Aoki, Y.; Mizuguchi, Y. SnAs-Based Layered Superconductor $NaSn_2As_2$. *J. Phys. Soc. Jpn.* **2017**, *86*, 123701.

(29) Sheldrick, G. M. Crystal Structure Refinement with SHELXL. *Acta Cryst C* **2015**, *71* (1), 3–8.

(30) Sheldrick, G. M. A Short History of SHELX. *Acta. Cryst. A* **2008**, *64*, 112–122.

(31) Spek, A. L. Single-Crystal Structure Validation with the Program PLATON. *J Appl. Cryst.* **2003**, *36*, 7–13.

(32) Spek, A. L. PLATON SQUEEZE: A Tool for the Calculation of the Disordered Solvent Contribution to the Calculated Structure Factors. *Acta Cryst C* **2015**, *71*, 9–18.

(33) Rodríguez-Carvajal, J. Recent Advances in Magnetic Structure Determination by Neutron Powder Diffraction. *Physica B: Condensed Matter* **1993**, *192*, 55–69.

(34) Hafner, J. Ab-Initio Simulations of Materials Using VASP: Density-Functional Theory and Beyond. *J. Comput. Chem.* **2008**, *29*, 2044–2078.

(35) Perdew, J. P.; Burke, K.; Ernzerhof, M. Generalized Gradient Approximation Made Simple. *Phys. Rev. Lett.* **1996**, *77*, 3865–3868.

(36) Dahl, J. P.; Avery, J. *Local Density Approximations in Quantum Chemistry and Solid State Physics*; Springer Science & Business Media, 2013.

(37) Anisimov, V. I.; Aryasetiawan, F.; Lichtenstein, A. I. First-Principles Calculations of the Electronic Structure and Spectra of Strongly Correlated Systems: The LDA+U method. *J. Phys.: Condens. Matter* **1997**, *9*, 767–808.

(38) Shim, J. H.; Haule, K.; Kotliar, G. Modeling the Localized-to-Itinerant Electronic Transition in the Heavy Fermion System CeIrIn$_5$. *Science* **2007**, *318*, 1615–1617.

(39) Monkhorst, H. J.; Pack, J. D. Special Points for Brillouin-Zone Integrations. *Phys. Rev. B* **1976**, *13*, 5188–5192.

(40) Kittel, C. *Introduction to Solid State Physics Global Edition*; Wiley: Hoboken, NJ, 2018.

(41) Ashcroft, N. W.; Mermin, N. D. *Solid State Physics*; Saunders College, 1976.

(42) Pletikosić, I.; Ali, M. N.; Fedorov, A. V.; Cava, R. J.; Valla, T. Electronic Structure Basis for the Extraordinary Magnetoresistance in WTe$_2$. *Phys. Rev. Lett.* **2014**, *113* (21), 216601.

(43) Ali, M. N.; Xiong, J.; Flynn, S.; Tao, J.; Gibson, Q. D.; Schoop, L. M.; Liang, T.; Haldolaarachchige, N.; Hirschberger, M.; Ong, N. P.; et al. Large, Non-Saturating Magnetoresistance in WTe$_2$. *Nature* **2014**, *514*, 205–208.

(44) Mong, R. S.; Essin, A. M.; Moore, J. E. Antiferromagnetic topological insulators. *Phys. Rev. B* **2010**, *81*, 245209.

(45) Fang, C.; Gilbert, M. J.; Bernevig, B. A. Topological Insulators with Commensurate Antiferromagnetism. *Phys. Rev. B* **2013** *88*, 085406.

(46) Yu, R.; Qi, X. L.; Bernevig, A.; Fang, Z.; Dai, X. Equivalent expression of Z$_2$ topological invariant for band insulators using the non-Abelian Berry connection, *Phys. Rev. B* **2011**, *84*, 075119.

(47) Skriver, H. L.; Rosengaard, N. M. Self-Consistent Green's-Function Technique for Surfaces and Interfaces. *Phys. Rev. B* **1991**, *43*, 9538–9549.

(48) King, P. D. C.; Hatch, R. C.; Bianchi, M.; Ovsyannikov, R.; Lupulescu, C.; Landolt, G.; Slomski, B.; Dil, J. H.; Guan, D.; Mi, J. L.; et al. Large Tunable Rashba Spin Splitting of a Two-Dimensional Electron Gas in Bi$_2$Se$_3$. *Phys. Rev. Lett.* **2011**, *107* (9), 096802.

**Table 1.** Single crystal structure refinement for EuSn$_2$P$_2$ at 293 (2) K.

| Refined Formula | EuSn$_2$P$_2$ |
| --- | --- |
| F.W. (g/mol) | 451.28 |
| Space group; Z | R -3m; 3 |
| $a$(Å) | 4.097 (1) |
| $c$(Å) | 26.162 (5) |
| V (Å$^3$) | 380.24 (13) |
| Extinction Coefficient | 0.0000 (1) |
| θ range (º) | 2.335-33.171 |
| No. reflections; $R_{int}$ | 2229; 0.0191 |
| No. independent reflections | 227 |
| No. parameters | 10 |
| $R_1$: $\omega R_2$ ($I>2\delta(I)$) | 0.0154; 0.0363 |
| Goodness of fit | 1.121 |
| Diffraction peak and hole (e$^-$/ Å$^3$) | 2.481; -0.617 |

**Table 2.** Atomic coordinates and equivalent isotropic displacement parameters for EuSn$_2$P$_2$ at 293 K. (U$_{eq}$ is defined as one-third of the trace of the orthogonalized U$_{ij}$ tensor (Å$^2$))

| Atom | Wyck. | Occ. | $x$ | $y$ | $z$ | $U_{eq}$ |
| --- | --- | --- | --- | --- | --- | --- |
| Eu1 | 3$a$ | 1 | 0 | 0 | 0 | 0.0088 (1) |
| Sn2 | 6$c$ | 1 | 0 | 0 | 0.2112 (1) | 0.0117 (1) |
| P3 | 6$c$ | 1 | 0 | 0 | 0.4053 (1) | 0.0087 (3) |

**Fig.1 The crystal structure of EuSn$_2$P$_2$ from different viewpoints, where the green, grey and red balls represent Eu, Sn and P atoms respectively.** **(A)** The single crystal picture of EuSn$_2$P$_2$. **(B) & (C)** The layered characteristic structure of EuSn$_2$P$_2$ (looking perpendicular to the *c*-axis.) **(D)** Emphasizing the honeycomb character of the double Sn layers in the basal plane and the triangular planes of Eu. (looking parallel to the *c*-axis.)

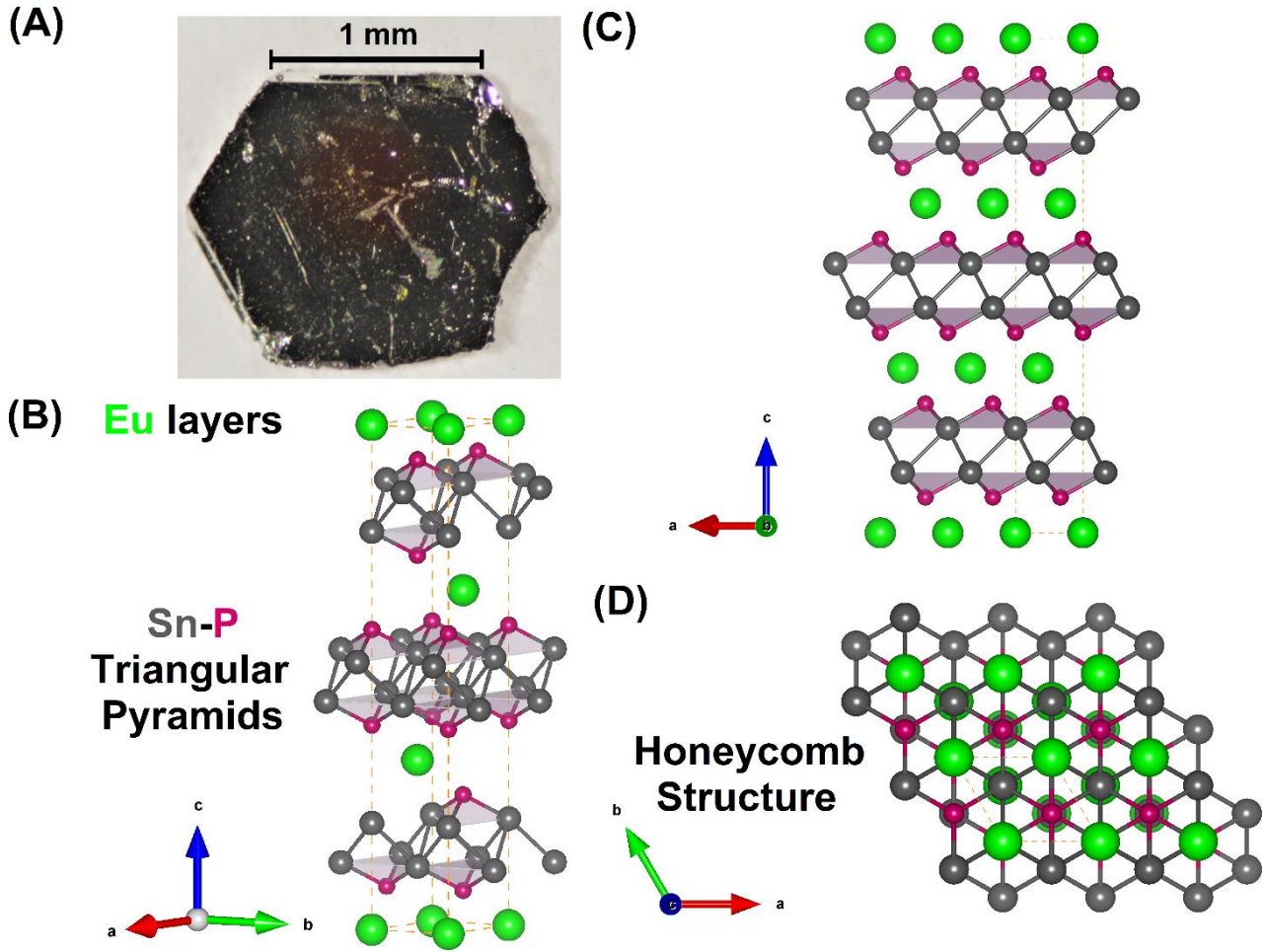

**Fig. 2 The Field-dependent, direction-dependent, and temperature-dependent magnetic properties of EuSn$_2$P$_2$.** **(A)** The field-dependent magnetization for the two crystal aorientations measured at 2 K where red balls and blue squares represent magnetic field parallel and perpendicular to the *c*-axis, respectively. **(B)** Magnetic susceptibility as a function of temperature measured in an external field of $\mu_0 H$ = 500 Oe, both perpendicular (blue) and parallel (red) to the *c*-axis. The inset shows the temperature dependent inverse susceptibility, measured for different sample orientations. The green line is the Curie-Weiss fit to the high-temperature data.

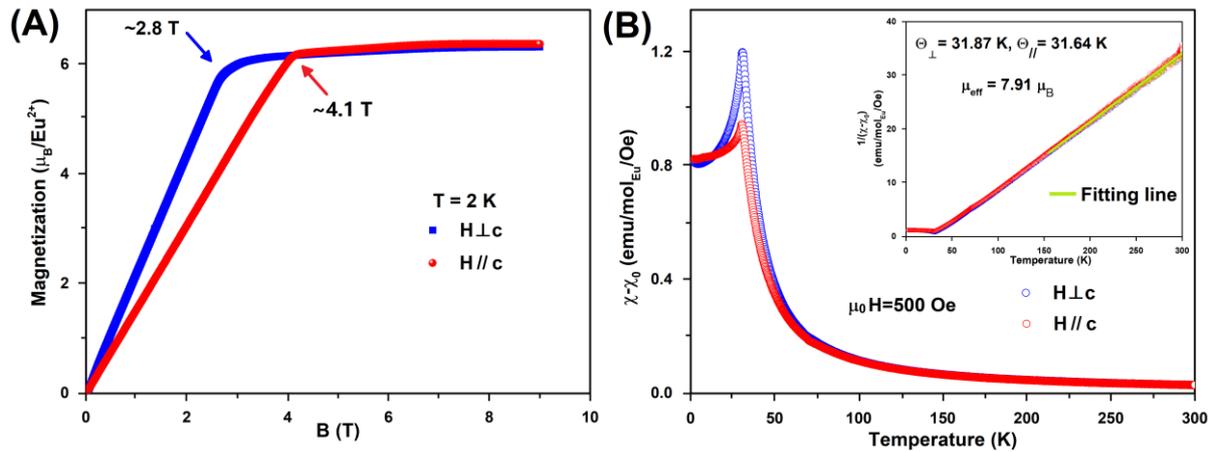

**Fig. 3 The characterization of the Eu spin state at ambient temperature and pressure in EuSn₂P₂.** Room temperature Electron paramagnetic resonance (EPR) measurement with applied magnetic field from 2300 to 5200 Gauss.

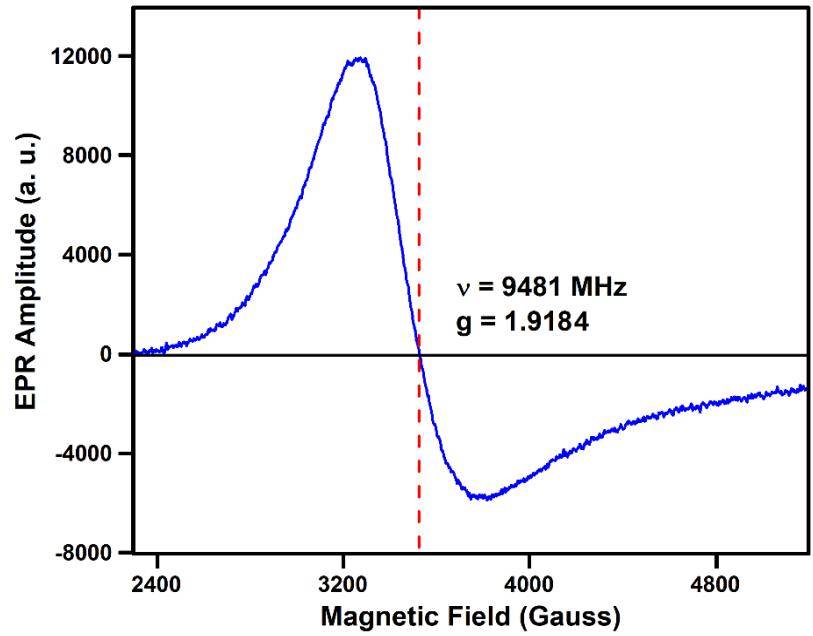

**Fig. 4 The Magnetic Structure of EuSn₂P₂.** **(A) (main panel)** The temperature dependence of the magnetic scattering. The pixel number is for the selected region of interest (enclosed in the purple rectangle inserts). The data indicates that $T_N \sim 30$ K, resulting from the fit of the red curves to the data. **(Inserts - Top right)** The magnetic signal intensity map at 30.0 K for the reciprocal lattice plane (0 0 4.5). **(Inserts bottom left)** The magnetic signal intensity map at 4.5 K for the reciprocal lattice plane (0 0 4.5). **(B)** The magnetic unit cell of EuSn₂P₂. The red arrows indicate the orientation of magnetic moments for each layer of Eu atoms.

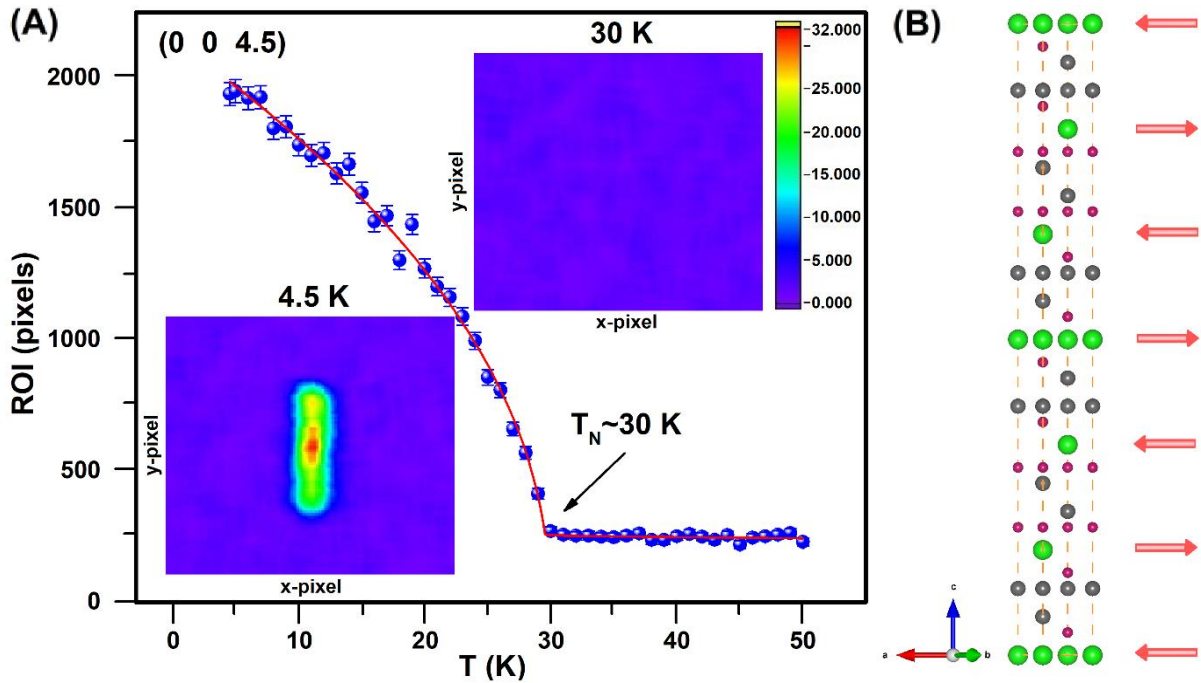

**Fig. 5 The charge transport properties of EuSn$_2$P$_2$** **(A)** the temperature dependence of the (001) plane resistivity $\rho_{xx}$ between 1.8 K and 300 K in the absence of an applied magnetic field. **(B)** Field-dependence of the Hall resistivity $\rho_{yx}$ at various temperatures (2 K, 10 K, 25 K, 70 K, 100 K and 150 K) with applied field from 0 to 9 Tesla. **(C)** The field-dependence of $\rho_{xx}$ for various temperatures (2 K, 5 K, 10 K, 15 K, 25 K, 40 K & 70 K) on sweeping applied field from 9 T to -9 T. **(D)** The magnetoresistance at the above temperatures normalized to the values at 0 applied magnetic field.

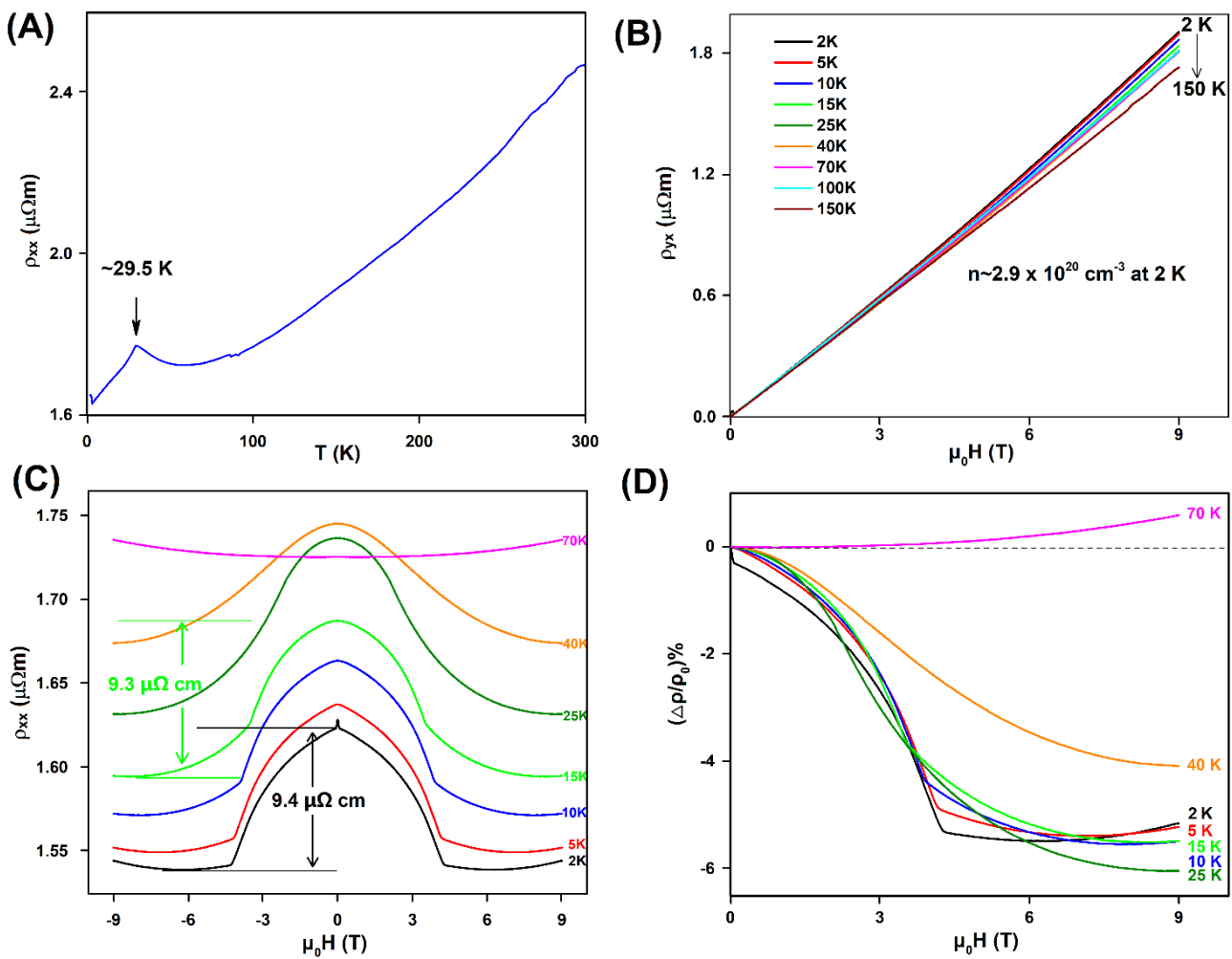

**Fig. 6 The calculated electronic structure of bulk EuSn$_2$P$_2$.** First-principles electronic band structures (A-D) for EuSn$_2$P$_2$ in its A-type antiferromagnetic ordered state (both spin up and spin down bands are shown) with and without spin-orbit coupling (SOC) and correlation parameter U included. Panel E shows the deconvolution of the electronic structure to show the energy regimes of dominance of the Eu, P, and Sn orbitals.

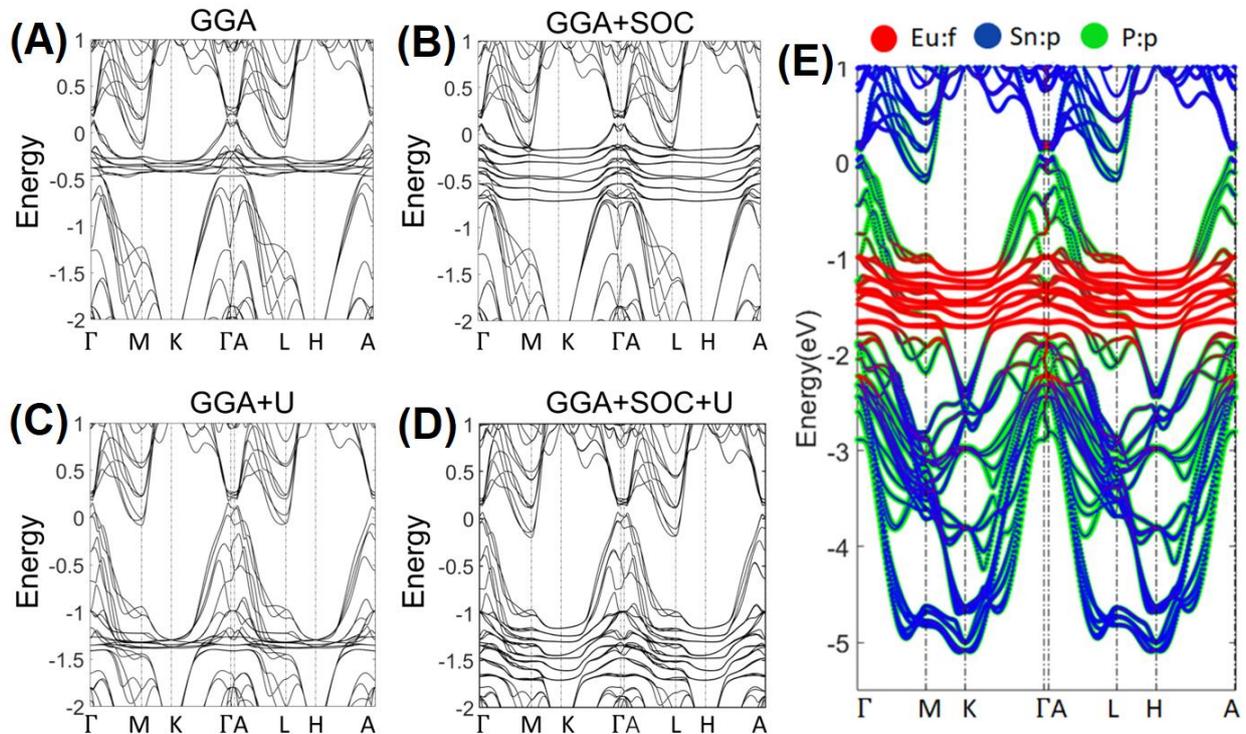

**Fig. 7 Comparison of the calculated bulk and surface band structures for EuSn$_2$P$_2$.** The bulk (left) and Surface (right) Band spectra of a semi-infinite slab with an (001) surface terminated at Sn (upper), P (middle) or Eu (lower) layers. The surface states are the narrow extra states observed in the surface state calculations (right).

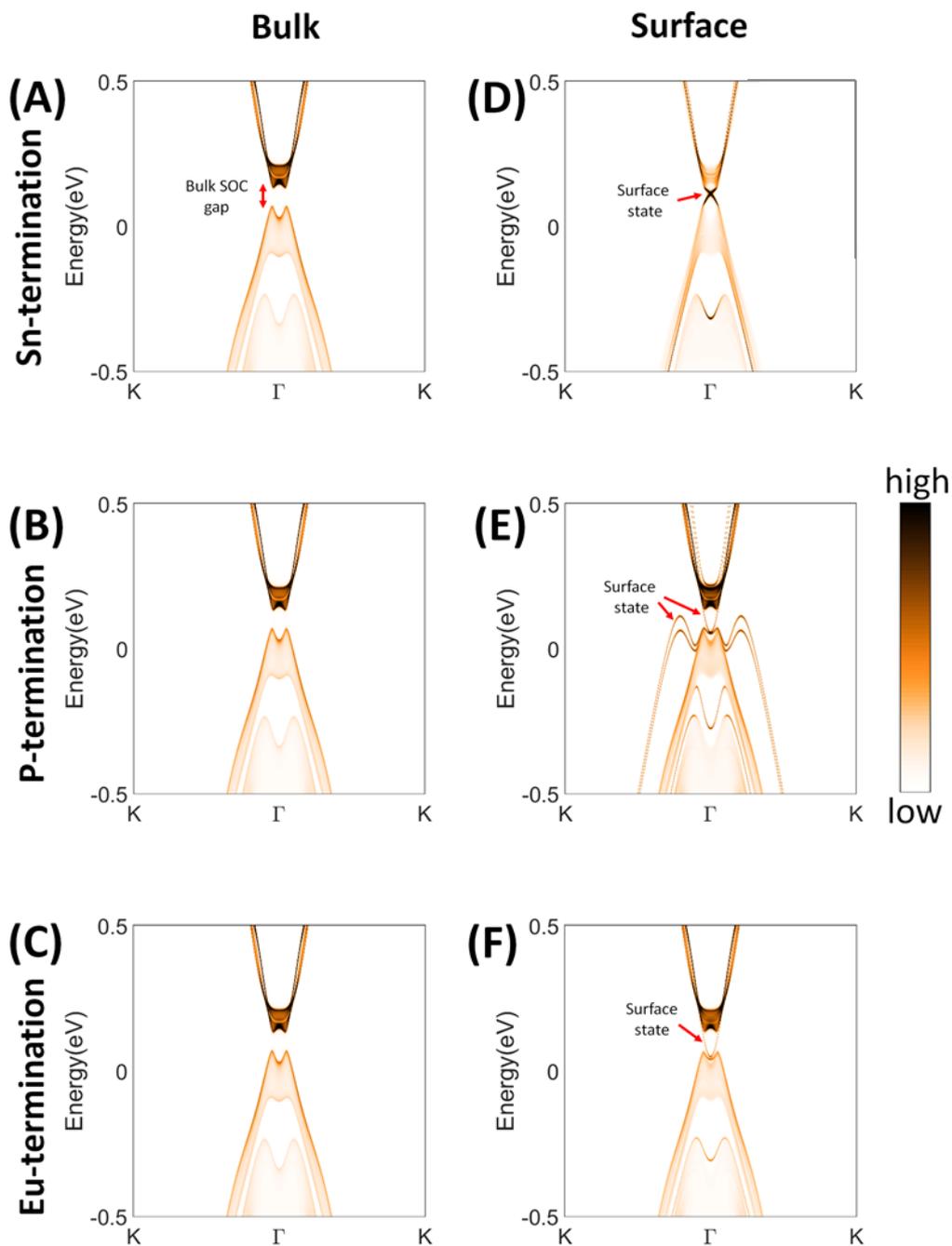

**Fig. 8 Topological invariant of EuSn$_2$P$_2$ (with SOC + U) with Wannier charge center evolution in time-reversal invariant planes.**

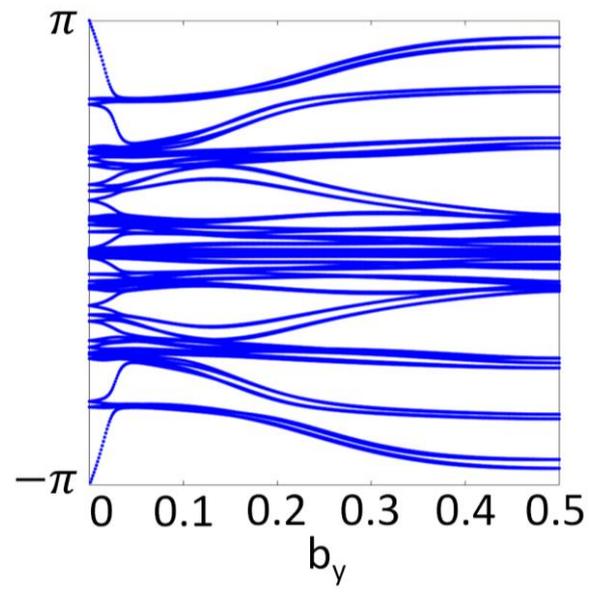

**Fig. 9 The calculated surface Fermi surfaces of EuSn$_2$P$_2$.** For the (001) surface (the cleavage plane), terminated by Sn, P, or Eu layers. The calculated iso-energy band contours are shown for E = 0 (the calculated Fermi Energy), -0.2, and -0.3 eV for Green's function surface spectral weight simulation. The depth of the color indicates the weight of charge density of each state at the surface layer. The surface states clearly appear as the extra ring-like Fermi surfaces in the panels.

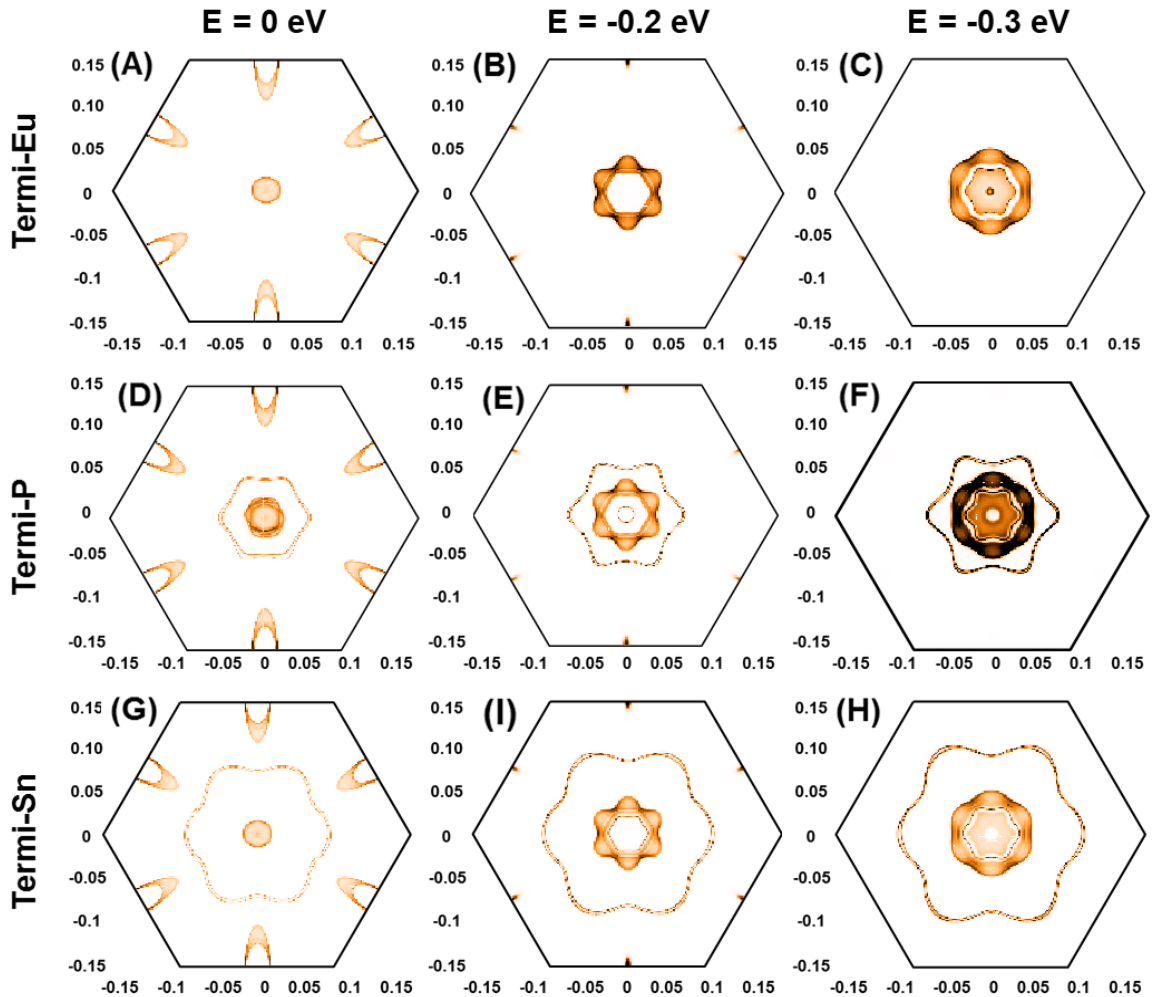

**Fig. 10 Angle-resolved photoemission study of the band structure of EuSn₂P₂. (A,B)** Band dispersion spectra along K-Γ-K taken with 76 and 96 eV photons differ mostly in the relative intensities of the bands due to photoemission selection-rule effects. Panel **C**, incident photon energy 102 eV, shows a wide energy-range band dispersion spectrum along M-Γ-M.

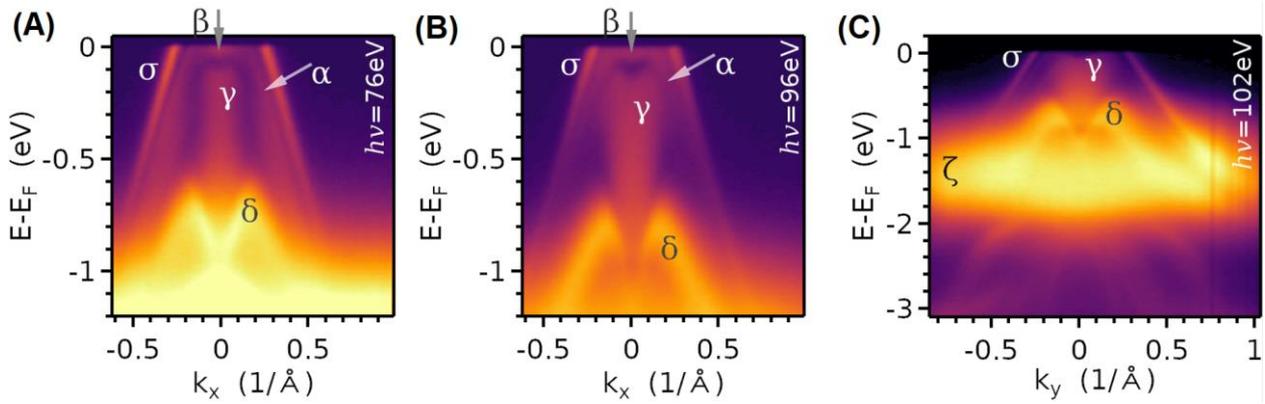

**Fig. 11 ARPES study continued.** **(A)** Two-dimensional momentum maps at several energies near the Fermi level obtained with two photon energies to enhance the visibility of different features in the band structure. The surface band, designated as σ, is clearly observed. **(B)** Photon energy-dependent map to follow the band dispersion along the *c* crystal axis. The plot is at the energy of -200 meV from $E_F$. ($V_0$ for the $k_z$ calculation was set to -10 eV). Different bands are marked with different Greek letters.

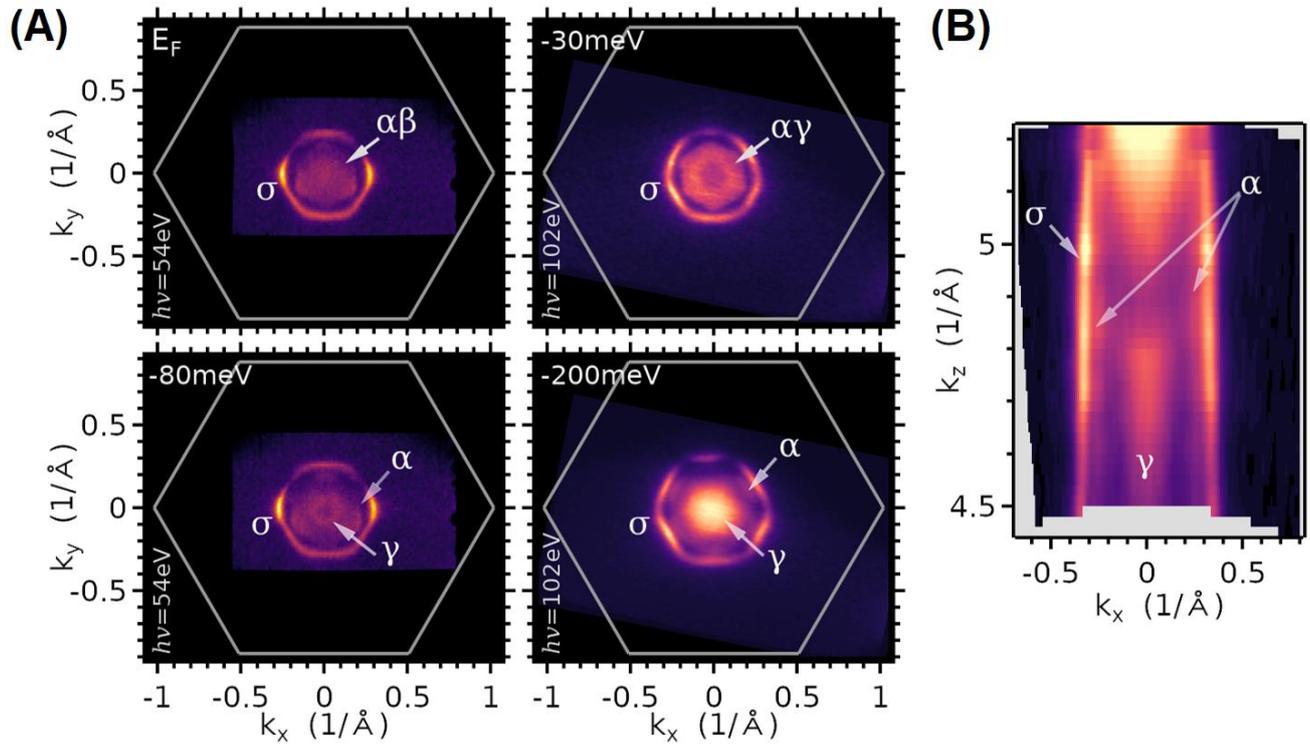

**Fig. 12 Chemistry of the surface** Core level spectrum of EuSn$_2$P$_2$ measured using 300 eV incident photons. This shows that the P has two different environments (there is only one type of P in the bulk crystal structure) and therefore is present in both the bulk and the surface.

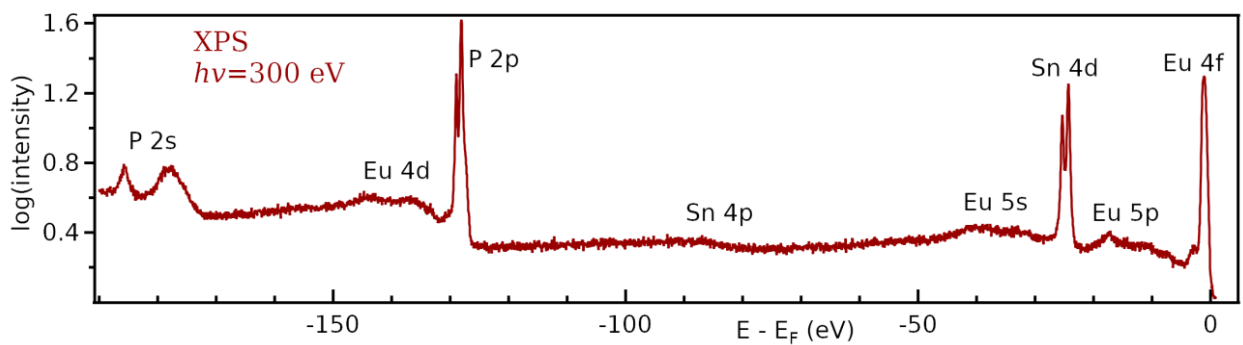